\documentclass{pj}
\usepackage{graphicx}

\begin{document}
\setcounter{page}{1}
\jheader{Vol.\ x, y--z, 2011}

\title{A General Relation Between Real and Imaginary Parts of the Magnetic Susceptibility}

\author{W.~G.~Fano and S.~Boggi and A.~C.~Razzitte}

\address{Faculty of Engineering, University of Buenos Aires\\
Paseo Colon Ave. 850, Cap. Fed. 1063, Argentina}  


\runningauthor{Fano}
\tocauthor{W.~G.}

\begin{abstract}
This paper is devoted to the study and the obtaining of the general relation between the real part and the imaginary part of the magnetic susceptibility function in the  Laplace domain.

This new theoretical technique is general, and can be applied to any magnetic material, that can be considered like causal and Linear time invariant (LTI).

A discussion of the causality which is extensively used in Physics has been done. To obtain the relations, some important concepts like Titchmarsh's theorem and Cauchy's Theorem have been reviewed, which results in the integral of a analytic function, that is formed with the magnetic susceptibility used in the Laplace domain.

The Cauchy Integral expression in the Laplace domain under certain conditions leads to a general relations between real and imaginary part of the magnetic susceptibility in the complex \textit{s}-plane.
These new relationships allow the validation of the magnetic susceptibilility functions developed by different researchers, in the Laplace domain, not just the frequency 
response like the well known Kramers-Kronig relations. Under certain conditions in these new general relations, the well known K-K relations can be obtained as a particular case.

 
\end{abstract}

%
\tableofcontents




\section{Introduction}
\label{Introduction}


It has been widely acknowledged that causality in a linear system and the K-K relations are equivalent in a way that they are necessary and sufficient conditions of each other, this can be seen in \cite{Wang}, \cite{Landau}, \cite{Toll}.

E. C. Titchmarsh (1926) has enounced a Theorem in the frequency complex plane, which shows the equivalence between causality and K-K relations, using Fourier Transform \cite{Nussenzveig}:
  
If square integrable function $G(\omega)$ fulfills one of the four conditions 
below, then it fulfills all four of them.
\begin{enumerate}
\item The inverse Fourier transform $g(t)$ of $G(\omega)$ vanishes:
$g(t)=0$ if $t<0$
\item $G(u)$ is for almost all $u$, the limit as $v\rightarrow 0^{+}$ of an analitic function $G(u+jv)$ that is holomorfic in the upper half plane and square integrable over any line parallel to the real axis:
$ \int _{-\infty} ^{+\infty} \mid G(u+jv)\mid ^2 du <C $ $v>0$
\item Re(G) and Im(G) verify the first Plemelj Formula:
$Re G(\omega)= \frac{1}{\pi}PV\int _{-\infty} ^{+\infty} \frac{Im(G(\omega'))}{\omega^{'}-\omega} d\omega' $
\item Re(G) and Im(G) verify the second Plemelj Formula:
$Im G(\omega)= \frac{1}{\pi}PV\int _{-\infty} ^{+\infty} \frac{Re(G(\omega'))}{\omega^{'}-\omega} d\omega'$
\end{enumerate}

where a function $f(x)$ is square integrable if $\int_{-\infty}^{\infty}|f(x)|^2dx$
is finite.

\vspace{.5cm}
A function $G(\omega)$ verifying one of the conditions of the Titchmarsh Theorem (and consequently all four of them), will be called causal transform. 
\vspace{.5cm}




In NiZn, MnZn, Ni2Y, and NiZnCu ferrites and their composites the causality and the numerical response have been investigated, using K-K relations \cite{Fano1}, where is possible to measure the real component of the complex magnetic permeability, in order to obtain the imaginary part of the magnetic permeability.

Usually the magnetic susceptibility of the MnZn and NiZn soft ferrites have been computed in the frequency domain \cite{Tsutaoka}, but another more powerfull way of analysis of the magnetic susceptibility function is by mean of the Laplace transform, wich has been recently obtained \cite{Fano20111708}. This Transformation allows the analysis of the system like the bounded-input, bounded-output (BIBO), and the application of the Routh-Hurwitz stability criterion.

The ferrite media under study can be considered as a linear, time invariant, isotropic and homogeneous, then the magnetization vector can be expressed by mean of a convolution product \cite{Fano20111708}, \cite{Geyi},\cite{Bladel}, thus:

\begin{equation}
 \overrightarrow{M}(t)=(\chi\ast\overrightarrow{H})(t)
\label{M(t)}
\end{equation}

where: $t$ is the time, and $\chi$ is the magnetic susceptibility.

Remembering the Laplace transform of $f(t)$ \cite{Wunch}:

\begin{equation}
L(f(t))=F(s)=\int_0 ^{\infty} e^{-st}f(t)dt
\end{equation}

Applying the Laplace transform to the magnetization vector $\overrightarrow{M}(t)$ of eqn. (\ref{M(t)}):

\begin{equation}
 \overrightarrow{M}(s)=\chi(s)\overrightarrow{H}(s)
 \label{Magnetization}
\end{equation}

where  $\chi(s)=\chi(s)^{'}-j\chi(s)^ {''}$ is the complex magnetic susceptibility in the Laplace domain.

Frequently the Fourier Transform is used to obtain the connection between the real and imaginary part of the magnetic susceptibility \cite{Landau}. In this paper the Laplace transformation is used to obtain these relations (see Appendix \ref{Kramers Kronig Relations with Laplace Transformation.}). 

\begin{equation}
\begin{array}{c}
\chi^{"}(\omega)=\frac{1}{\pi}PV \int _{-\infty} ^{\infty} \left(\frac{\chi^{'}(\xi)}{\xi-\omega}\right)d\xi\\
\chi^{'}(\omega)=-\frac{1}{\pi} PV \int _{-\infty} ^{\infty} \left(\frac{\chi^{"}(\xi)}{\xi-\omega}\right)d\xi
\label{Relaciones de K K}
\end{array}
\end{equation}

where the integral is the well known Hilbert transform \cite{Landau}, \cite{Fano1}.

\section{Formulation}
\begin{figure}[ht]
\centerline{\includegraphics[width=0.35\columnwidth,draft=false]{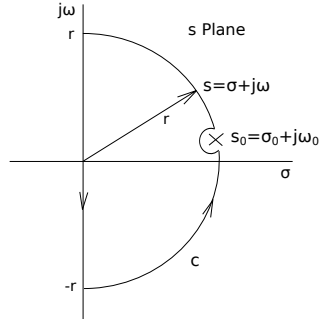}}
\caption{Integration's path "C" in the complex s-plane}
\label{fig:fig5_2011}
\end{figure}

A general relation between real and imaginary part of the magnetic susceptibility in the complex \textit{s}-plane will be developed in this paper. This will be obtained by mean of the same integral equation used in K-K of Appendix \ref{Kramers Kronig Relations with Laplace Transformation.}, with the singularity placed on the arc as is depicted in the Figure \ref{fig:fig5_2011}:

\begin{equation}
\oint _c \left(\frac{\chi(s)}{s-s_0}\right)ds=0
\end{equation}

where $\chi(s)$ is the magnetic susceptibility function in the Laplace domain, that is analytic in the half right of the complex \textit{s}-plane.

The integral can be expressed as a sum of two terms:

\begin{equation}
\oint _c \left(\frac{\chi(s)}{s-s_0}\right)ds=
\oint _{Ca} \left(\frac{\chi(s)}{s-s_0}\right)ds+
\oint _{Cb} \left(\frac{\chi(s)}{s-s_0}\right)ds
\end{equation}

\subsection{Integral over "Ca"}

\begin{figure}[ht]
\centerline{\includegraphics[width=0.4\columnwidth,draft=false]{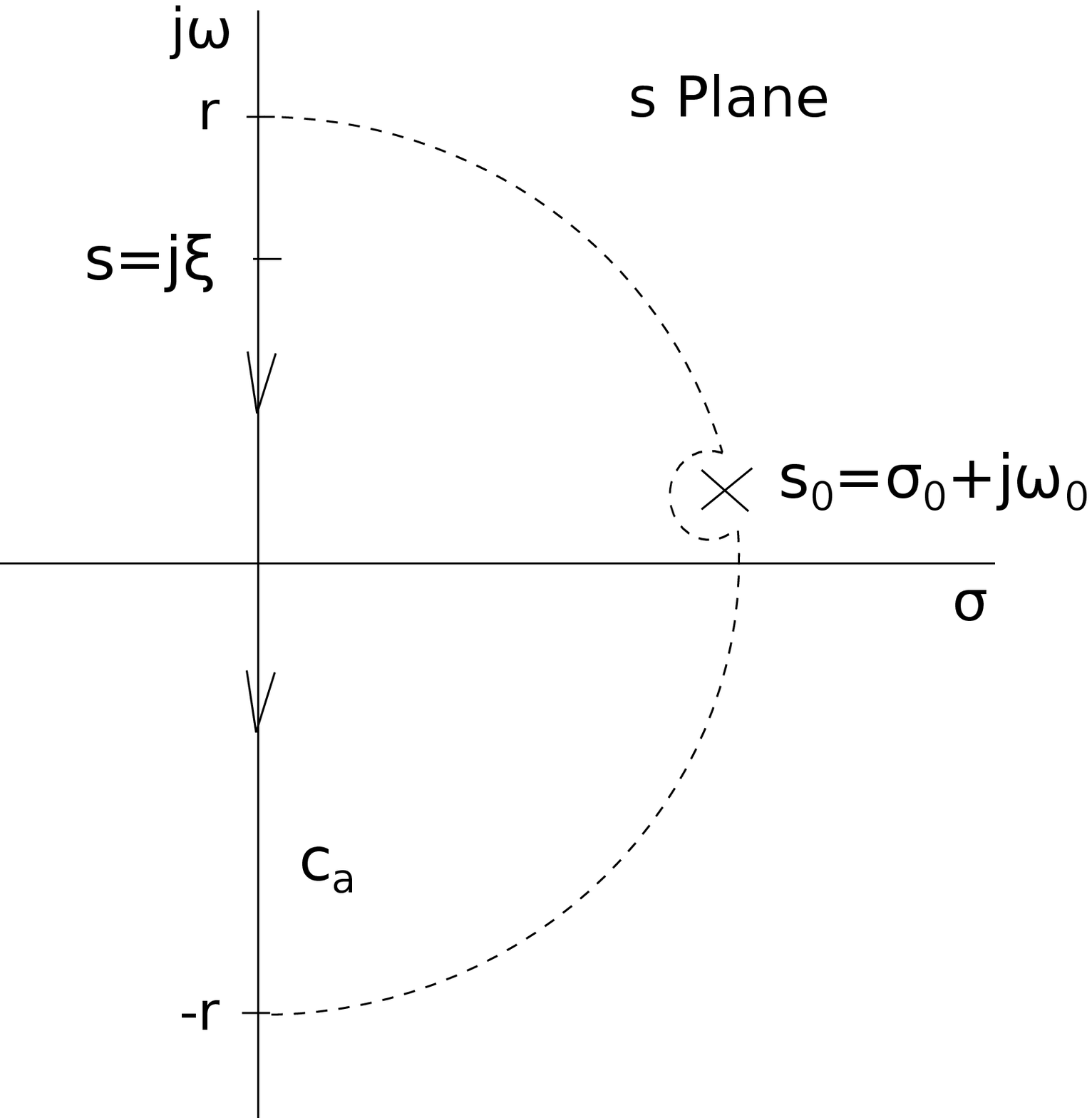}}
\caption{Integration's path "Ca" in the complex s-plane}
\label{fig:fig6_2011}
\end{figure}

\begin{equation}
\oint _{Ca} \left(\frac{\chi(s)}{s-s_0}\right)ds
\end{equation}

Assuming:

\begin{equation}
s_0=\sigma_0+j\omega_0
\end{equation}

\begin{equation}
s=j\xi
\end{equation}

\begin{equation}
s-s_0=j\xi-(\sigma_0+j\omega_0)=-\sigma_0+j(\xi-\omega_0)
\end{equation}

The integral can be expressed thus:

\begin{equation}
\int _{-r} ^{+r} \left(\frac{\chi(s)}{s-s_0}\right)ds
\end{equation}

where: $s=j\xi$

\begin{equation}
\int _{-r} ^{+r} \left(\frac{\chi(j\xi)}{-\sigma_0+j(\xi-\omega_0)}\right)jd\xi
\end{equation}

Resolving:

\begin{equation}
j\int _{-r} ^{+r} \left(\frac{\chi(j\xi)}{-\sigma_0+j(\xi-\omega_0)}\right)
\left(\frac{-\sigma_0-j(\xi-\omega_0)}{-\sigma_0-j(\xi-\omega_0)}\right)d\xi
\end{equation}

Then:

\begin{equation}
j\int _{-r} ^{+r} \left(\frac{\chi(j\xi)}{\sigma^2_0+(\xi-\omega_0)^2}\right)
\left(-\sigma_0-j(\xi-\omega_0)\right)d\xi
\end{equation}

Using: 

\begin{equation}
\chi(\xi)=\chi^{'}(\xi)-j\chi^{"}(\xi)
\end{equation}

Then:

\begin{equation}
\int _{-r} ^{+r} \left(\frac{\left(\chi^{'}(\xi)\right)(\xi-\omega_0)
-\sigma_0 \left(\chi^{"}(\xi)\right)}{\sigma^2_0+(\xi-\omega_0)^2}\right)
d\xi
\end{equation}

\begin{equation}
-j\int _{-r} ^{+r} \left(\frac{\left(\chi^{'}(\xi)\right)\sigma_0
+\left(\chi^{"}(\xi)\right)(\xi-\omega_0) }{\sigma^2_0+(\xi-\omega_0)^2}\right)
d\xi
\end{equation}

\subsection{Integral on the path "Cb"}

\begin{figure}[ht]
\centerline{\includegraphics[width=0.4\columnwidth,draft=false]{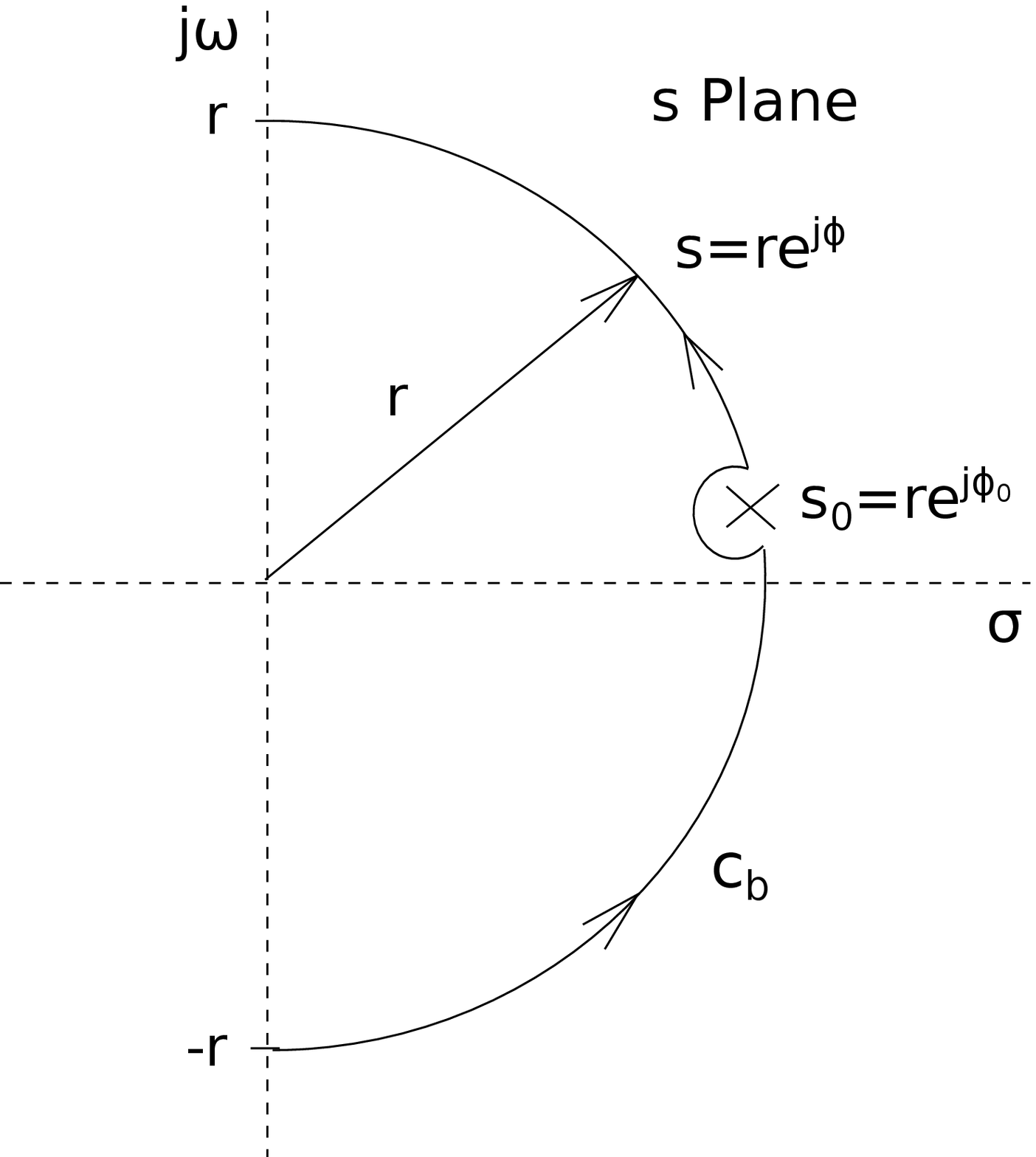}}
\caption{Integration's path "Cb" in the complex s-plane}
\label{fig:fig7_2011}
\end{figure}

\begin{equation}
\int _{Cb} \left(\frac{\chi(s)}{s-s_0}\right)ds
\end{equation}

The integral can be expressed thus:

\begin{equation}
\begin{array}{c}
\int _{C} \left(\frac{\chi(s)}{s-s_0}\right)ds=\int _{C_1} \left(\frac{\chi(s)}{s-s_0}\right)ds\\
+\int _{C_2}\left(\frac{\chi(s)}{s-s_0}\right)ds+\int _{C_3} \left(\frac{\chi(s)}{s-s_0}\right)ds 
\end{array}
\end{equation}

where $C_2$ is the path around the singularity. $C_1$ and $C_3$ are the rest of the arc

Assuming:

\begin{equation}
s_0=re^{j\phi_0}
\end{equation}

\begin{equation}
s=re^{j\phi}
\end{equation}

Then:

\begin{equation}
s-s_0=re^{j\phi}-re^{j\phi_0}
\end{equation}

\begin{equation}
ds=re^{j\phi}d\phi
\end{equation}

The integral is:

\begin{equation}
\begin{array}{c}
\int _{C_1} ^{} \left(\frac{\chi(s)}{re^{j\phi}-re^{j\phi_0}}\right)re^{j\phi}d\phi\\
-j\pi\left(\chi(s_0)\right)\\
+\int _{C_3} \left(\frac{\chi(s)}{re^{j\phi}-re^{j\phi_0}}\right)re^{j\phi}d\phi
\end{array}
\end{equation}

Solving:

\begin{equation}
\begin{array}{c}
\int _{C_1} \left(\frac{\chi(s)}{1-e^{j(\phi_0-\phi)}}\right)d\phi\\
-j\pi\left(\chi(s_0)\right)\\
+\int _{C_3} \left(\frac{\chi(s)}{1-e^{j(\phi_0-\phi)}}\right)d\phi
\end{array}
\end{equation}

Using:

\begin{equation}
1-e^{j(\phi_0-\phi)}= 1-cos(\phi_0-\phi)-jsen(\phi_0-\phi)
\end{equation}

\begin{equation}
\frac{1}{e^{j(\phi_0-\phi)}}=
\frac{1-cos(\phi_0-\phi)+jsen(\phi_0-\phi)}{(1-cos(\phi_0-\phi))^2+sen^2(\phi_0-\phi)}
\end{equation}

where the real and imaginary parts:

\begin{equation}
\frac{1}{e^{j(\phi_0-\phi)}}=
\frac{1-cos(\phi_0-\phi)}{(1-cos(\phi_0-\phi))^2+sen^2(\phi_0-\phi)}
+\frac{jsen(\phi_0-\phi)}{(1-cos(\phi_0-\phi))^2+sen^2(\phi_0-\phi)}
\end{equation}

assuming:

\begin{equation}
\frac{1}{e^{j(\phi_0-\phi)}}= K_1(\phi_0,\phi)+jK_2(\phi_0,\phi)
\end{equation}

\begin{equation}
\begin{array}{c}
\int _{0} ^{\phi_0-} \left( \chi^{'}(s)-j\chi^{"}(s)\right)(K_1(\phi_0,\phi)+jK_2(\phi_0,\phi))d\phi\\
-j\pi\left(\chi^{'}(s_0)-j\chi^{"}(s_0)\right)\\
+\int _{\phi_0+} ^{\pi}\left( \chi^{'}(s)-j\chi^{"}(s)\right)(K_1(\phi_0,\phi)+jK_2(\phi_0,\phi))d\phi
\end{array}
\end{equation}

If the radius $\rho$ of the path around the singularity $s_0$, $\rho\rightarrow 0$:

\begin{equation}
\begin{array}{c}
PV\int _{0} ^{\pi} \left( \chi^{'}(s)-j(\chi^{"}(s))\right)(K_1(\phi_0,\phi)+jK_2(\phi_0,\phi))d\phi\\
-j\pi\left(\chi^{'}(s_0)-j\chi^{"}(s_0)\right)\\
\end{array}
\end{equation}

With the real and imaginary parts :

\begin{equation}
\begin{array}{c}
\pi\left(-\chi^{"}(s_0)\right)+PV\int _{0} ^{\pi} \left(\chi^{'}(s))\right)(K_1(\phi_0,\phi)+(\chi^{"}(s))K_2(\phi_0,\phi))d\phi\\
-j\pi\left(\chi^{'}(s_0)\right)+jPV\int _{0} ^{\pi} (\left( \chi^{'}(s)\right)K_2(\phi_0,\phi)
-(\chi^{"}(s))K_1(\phi_0,\phi))d\phi\\
\end{array}
\end{equation}

Rewriting the Cauchy theorem:

\begin{equation}
\oint _c \left(\frac{\chi(s)}{s-s_0}\right)ds=
\oint _{Ca} \left(\frac{\chi(s)}{s-s_0}\right)ds+
\oint _{Cb} \left(\frac{\chi(s)}{s-s_0}\right)ds=0+j0
\end{equation}

Then the real part and the imaginary part are:

\begin{equation}
\begin{array}{c}
-\pi\chi^{"}(s_0)+PV\int _{0} ^{\pi} (\chi^{'}(s)K_1(\phi_0,\phi)+\chi^{"}(s)K_2(\phi_0,\phi))d\phi\\
+\int _{-r} ^{+r} \frac{\chi^{'}(\xi)(\xi-\omega_0)
-\sigma_0 \chi^{"}(\xi)}{\sigma^2_0+(\xi-\omega_0)^2}
d\xi=0
\end{array}
\end{equation}

\begin{equation}
\begin{array}{c}
-j\pi\chi^{'}(s_0)+jPV\int _{0} ^{\pi} (\chi^{'}(s)K_2(\phi_0,\phi)
-\chi^{"}(s)K_1(\phi_0,\phi))d\phi\\
-j\int _{-r} ^{+r} \frac{\chi^{'}(\xi)\sigma_0
+\chi^{"}(\xi)(\xi-\omega_0) }{\sigma^2_0+(\xi-\omega_0)^2}
d\xi=0
\end{array}
\end{equation}

Finally the real part:

\begin{equation}
\begin{array}{c}
PV\int _{0} ^{\pi} \chi^{'}(s)K_1(\phi_0,\phi)d\phi+\int _{-r} ^{+r} \frac{\chi^{'}(\xi)(\xi-\omega_0)
}{\sigma^2_0+(\xi-\omega_0)^2}
d\xi\\
= \pi\chi^{"}(s_0)- PV\int _{0} ^{\pi} \chi^{"}(s)K_2(\phi_0,\phi)d\phi
+\int _{-r} ^{+r} \frac{\sigma_0 \chi^{"}(\xi)}{\sigma^2_0+(\xi-\omega_0)^2}d\xi
\end{array}
\label{chi real nueva}
\end{equation}

and the imaginary part:

\begin{equation}
\begin{array}{c}
-PV\int _{0} ^{\pi} 
\chi^{"}(s)K_1(\phi_0,\phi)d\phi
-\int _{-r} ^{+r} \frac{\chi^{"}(\xi)(\xi-\omega_0) }{\sigma^2_0+(\xi-\omega_0)^2}
d\xi\\
=\pi\chi^{'}(s_0)
-PV\int _{0} ^{\pi} \chi^{'}(s)K_2(\phi_0,\phi)d\phi
+\int _{-r} ^{+r}\frac{\chi^{'}(\xi)\sigma_0}{\sigma^2_0+(\xi-\omega_0)^2}
d\xi
\end{array}
\label{chi imag nueva}
\end{equation}


\section{Results and Discussion}
\subsection{Solving the equations}
The new general relations obtained, shows a connection between the real and imaginary part of the magnetic susceptibility. Rewritting the eqn. (\ref{chi real nueva}) that corresponds to the imaginary part:

\begin{equation}
\begin{array}{c}
PV\int _{0} ^{\pi} \chi^{'}(s)K_1(\phi_0,\phi)d\phi+\int _{-r} ^{+r} \frac{\chi^{'}(\xi)(\xi-\omega_0)}{\sigma^2_0+(\xi-\omega_0)^2}
d\xi\\
= \pi\chi^{"}(s_0)- PV\int _{0} ^{\pi} \chi^{"}(s)K_2(\phi_0,\phi)d\phi
+\int _{-r} ^{+r} \frac{\sigma_0 \chi^{"}(\xi)}{\sigma^2_0+(\xi-\omega_0)^2}d\xi
\end{array}
\label{chi real nueva2}
\end{equation}

Neglecting the integrals of the right side, the imaginary part of the magnetic susceptibility  $\chi^{"}$ can be calculated, as a first step. 

\begin{equation}
PV\int _{0} ^{\pi} \chi^{'}(s)K_1(\phi_0,\phi)d\phi+\int _{-r} ^{+r} \frac{\chi^{'}(\xi)(\xi-\omega_0)
}{\sigma^2_0+(\xi-\omega_0)^2}
d\xi= \pi\chi^{"}(s_0)
\end{equation}

After this, an iterative method can be used with the complete eqn.
(\ref{chi real nueva2}), but it is not employed in this paper.

\subsection{Obtaining the K-K relations from the general equations}
Another interesting test can be performed from the equations (\ref{chi real nueva}) and (\ref{chi imag nueva}).

\begin{equation}
\begin{array}{c}
PV\int _{0} ^{\pi} \chi^{'}(s)K_1(\phi_0,\phi)d\phi+\int _{-r} ^{+r} \frac{\chi^{'}(\xi)(\xi-\omega_0)
}{\sigma^2_0+(\xi-\omega_0)^2}
d\xi\\
= \pi\chi^{"}(s_0)- PV\int _{0} ^{\pi} \chi^{"}(s)K_2(\phi_0,\phi)d\phi
+\int _{-r} ^{+r} \frac{\sigma_0 \chi^{"}(\xi)}{\sigma^2_0+(\xi-\omega_0)^2}d\xi
\end{array}
\label{KK1}
\end{equation}

\begin{equation}
\begin{array}{c}
-PV\int _{0} ^{\pi} 
\chi^{"}(s)K_1(\phi_0,\phi)d\phi
-\int _{-r} ^{+r} \frac{\chi^{"}(\xi)(\xi-\omega_0) }{\sigma^2_0+(\xi-\omega_0)^2}
d\xi\\
=\pi\chi^{'}(s_0)
-PV\int _{0} ^{\pi} \chi^{'}(s)K_2(\phi_0,\phi)d\phi
+\int _{-r} ^{+r}\frac{\chi^{'}(\xi)\sigma_0}{\sigma^2_0+(\xi-\omega_0)^2}
d\xi
\end{array}
\label{KK2}
\end{equation}

If the radius of the arc are decresing, up to a coincidence between both path in the Figure \ref{fig:fig5_2011}, then: $\sigma_0=0$. In this situation $\frac{\partial }{\partial \phi}=0$. Then the equations (\ref{KK1}) and (\ref{KK2}) can be written thus: 

\begin{equation}
\begin{array}{c}
\chi^{"}(\omega)=\frac{1}{\pi}PV \int _{-r} ^{+r} \frac{\chi^{'}(\xi)}{\xi-\omega}d\xi\\
\chi^{'}(\omega)=-\frac{1}{\pi} PV \int _{-r} ^{+r} \frac{\chi^{"}(\xi)}{\xi-\omega}d\xi
\end{array}
\end{equation}

If $r\rightarrow\infty$ in order to cover the half right \textit{s}-complex plane. The well known K-K equations explained in \ref{Relaciones de K K} are obtained: 

\begin{equation}
\begin{array}{c}
\chi^{"}(\omega)=\frac{1}{\pi}PV \int _{-\infty} ^{\infty} \frac{\chi^{'}(\xi)}{\xi-\omega}d\xi\\
\chi^{'}(\omega)=-\frac{1}{\pi} PV \int _{-\infty} ^{\infty} \frac{\chi^{"}(\xi)}{\xi-\omega}d\xi
\end{array}
\end{equation}

The K-K relations are a particular case of these new general relations written in eqns. (\ref{chi real nueva}) and (\ref{chi imag nueva})

\section{Conclusions}

In this paper the magnetic susceptibility has been considered as a linear, time invariant, isotropic and homogeneous. The study of a new relation between the real and the imaginary part of the magnetic susceptibility function in the  Laplace domain, in the 
\textit{s}-complex plane have been realized. 

A discussion of the causality and another concepts like Titchmarsh's theorem and Cauchy's Theorem have been done, considering a analytic function, that is formed with the magnetic susceptibility in the Laplace domain.

By mean of the Cauchy theorem in the Laplace domain under certain conditions leads to a general relations between real and imaginary part of the magnetic susceptibility in the complex \textit{s}-plane.

The K-K relations can be obtained reducing these new general relations under certain assumptions, because K-K realtions are a particular case of these relations.

These new relationships allow the validation of the magnetic susceptibilility functions developed by different researchers, in the Laplace domain, not just the frequency 
response like the well known Kramers-Kronig relations. 

These new general relations could be applied to the dielectric materials as well. 












\appendixx{K-K Relations with Laplace transformation}
\label{Kramers Kronig Relations with Laplace Transformation.}
\begin{figure}[ht]
\centerline{\includegraphics[width=0.4\columnwidth,draft=false]{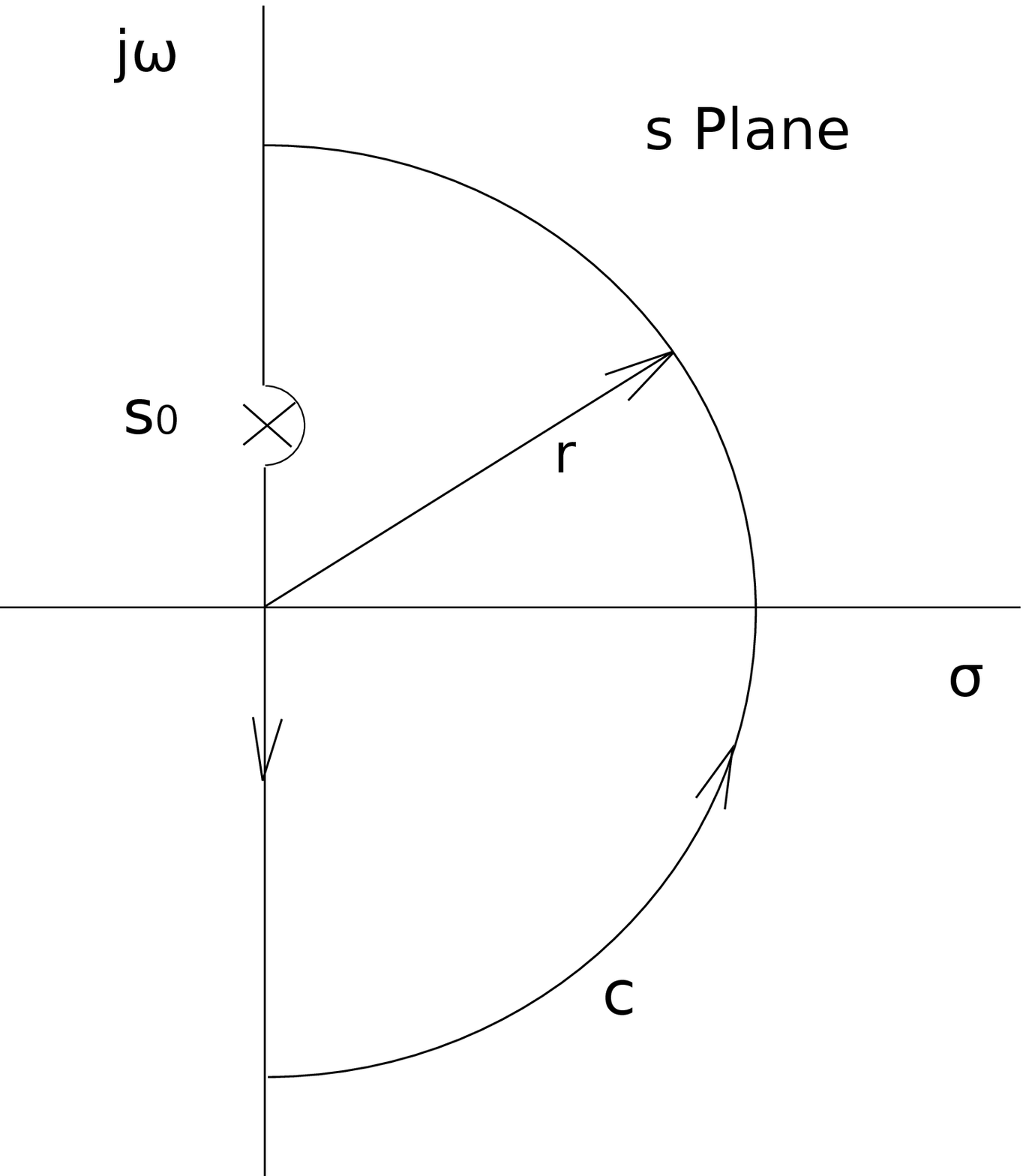}}
\caption{Integration's path "C" on the complex \textit{s}-plane to obtain K-K relations in the Laplace Domain}
\label{figApp:fig1_2011}
\end{figure}

Using the expresion of the magnetic susceptibility in the Laplace domain 
\cite{Fano20111708}, the well known K-K equation can be obtained from the integral over the path "c" on the complex \textit{s}-plane of the following function (see Figure \ref{figApp:fig1_2011}):

\begin{equation}
\oint _c \left(\frac{\chi(s)}{s-s_0}\right)ds
\label{eqn planteo}
\end{equation}

where 
$\chi(s)$ is the magnetic susceptibility in the Laplace domain 

The function $\left(\frac{\chi(s)}{s-s_0}\right)$ on the path c and inside the domain doesn't has any poles, and is analytic, then the Cauchy theorem can be used 
\cite{Wunch}:

\begin{equation}
\oint _c \left(\frac{\chi(s)}{s-s_0}\right)ds=0
\label{Cauchy}
\end{equation}

Using the path of the Figure \ref{figApp:fig1_2011}, the integral of eqn. (\ref{Cauchy})
can be expressed like a sum of two terms: one term is the circunference arc (Ca) and the other term is the imaginary axe $\omega$ (Cb). 

\subsection{Integral in (Ca)}

\begin{figure}[ht]
\centerline{\includegraphics[width=0.35\columnwidth,draft=false]{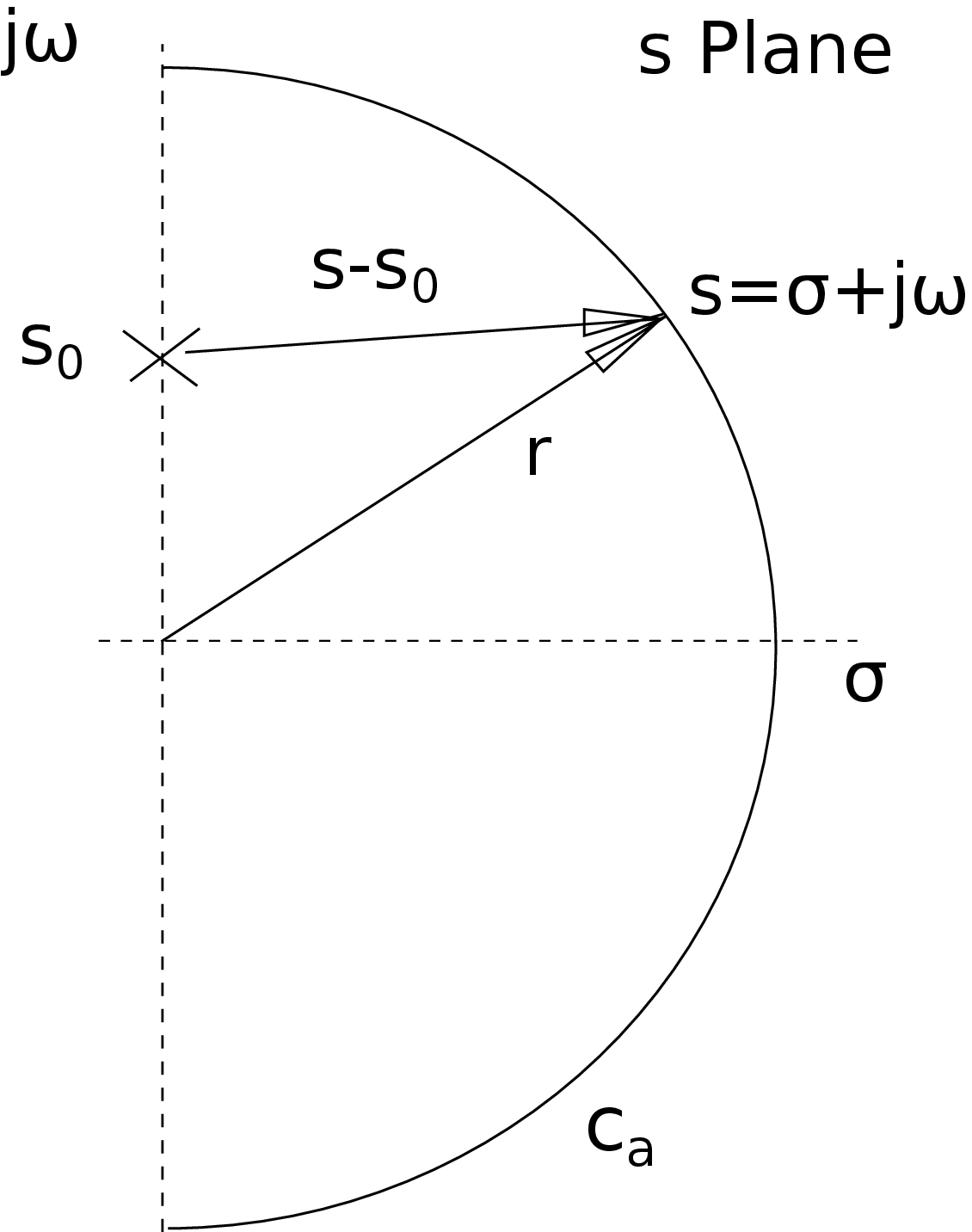}}
\caption{Integration's path "Ca" in the complex \textit{s}-plane of the arc of the circumference}
\label{figApp:fig4_2011}
\end{figure}

As the radio of the half circumference tends to infinity, then:
 
\begin{equation}
\lim_{r\to \infty} \left(\frac{\chi(s)}{s-s_0}\right)=0
\end{equation}

Results:

\begin{equation}
\int _{ca} \left(\frac{\chi(s)}{s-s_0}\right)ds=0
\label{Ca}
\end{equation}

\subsection{Integral in Cb}

\begin{figure}[ht]
\centerline{\includegraphics[width=0.35\columnwidth,draft=false]{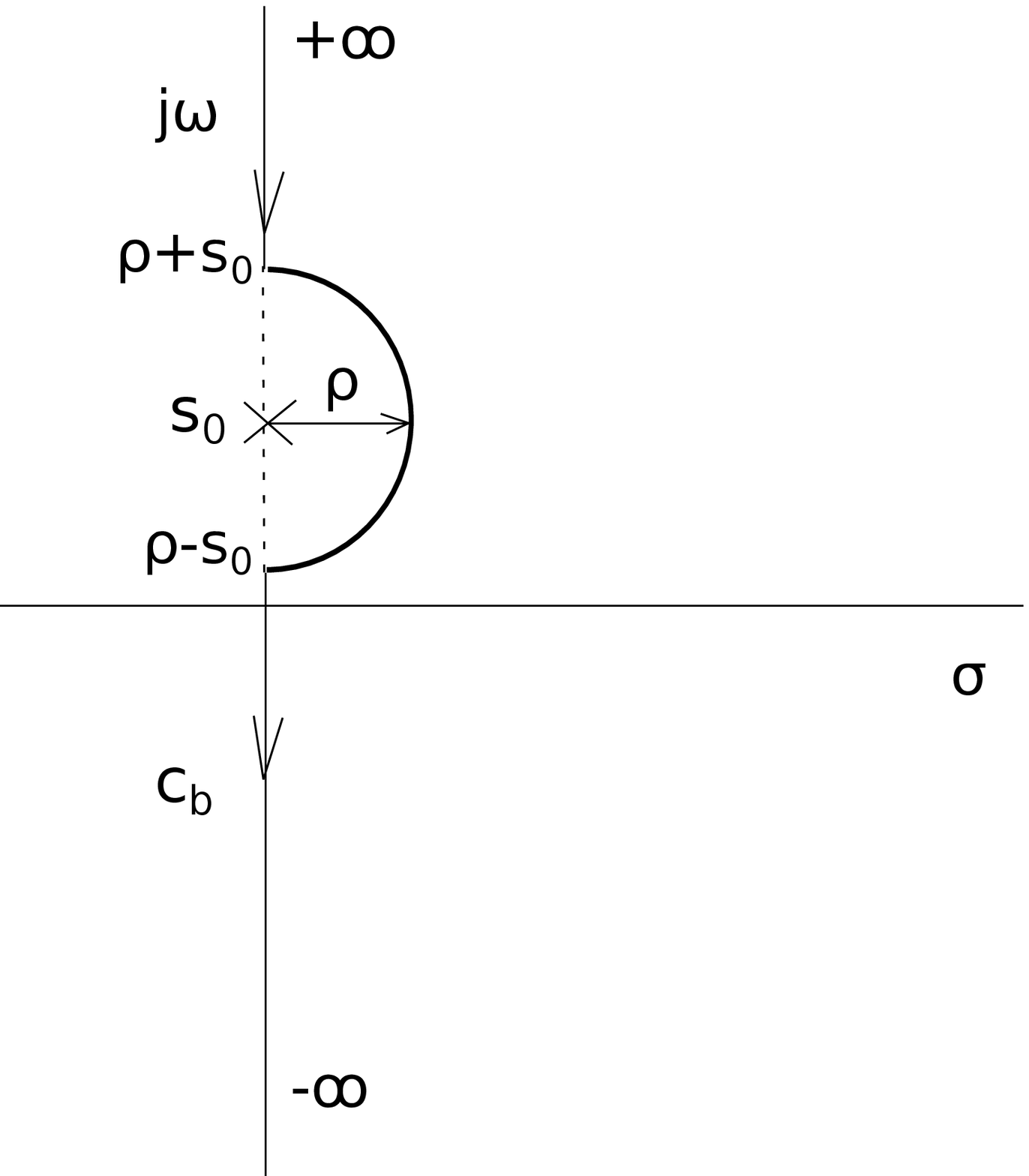}}
\caption{Integration's path "Cb" in the complex \textit{s}-plane of the imaginary axis}
\label{figApp:fig2_2011}
\end{figure}

The integral on the path Cb can be expressed in three terms, this can be observed in the Figure \ref{figApp:fig2_2011}:

\begin{equation}
\begin{array}{c}
\oint _{cb} \left(\frac{\chi(s)}{s-s_0}\right)ds=\int _{\infty} ^{\rho +s_0} \left(\frac{\chi(s)}{s-s_0}\right)ds\\
+\int _{\rho +s_0} ^{\rho -s_0} \left(\frac{\chi(s)}{s-s_0}\right)ds +\int _{\rho -s_0} ^{-\infty}  \left(\frac{\chi(s)}{s-s_0}\right)ds
\end{array}
\end{equation}

If $\rho \to 0$:

\begin{equation}
\begin{array}{c}
\oint _{cb} \left(\frac{\chi(s)}{s-s_0}\right)ds=
\lim_{\rho \to 0}\left(\int _{\rho +s_0} ^{\rho -s_0} \left(\frac{\chi(s)}{s-s_0}\right)ds \right)\\
+PV \int _{\infty} ^{-\infty} \left(\frac{\chi(s)}{s-s_0}\right)ds
\end{array}
\end{equation}

The contribution of the first integral is $-j\pi\chi(s_0)$ \cite{Landau}:

\begin{equation}
\oint _{cb} \left(\frac{\chi(s)}{s-s_0}\right)ds=-j\pi\chi(s_0)+ PV \int _{\infty} ^{-\infty} \left(\frac{\chi(s)}{s-s_0}\right)ds
\label{Cb}
\end{equation}

Using :

\begin{equation}
\oint _{ca+cb} \left(\frac{\chi(s)}{s-s_0}\right)ds=0
\end{equation}

Considering (\ref{Cb}), (\ref{Ca}), and (\ref{Cauchy}):

\begin{equation}
-j\pi\chi(s_0)+ PV \int _{\infty} ^{-\infty} \left(\frac{\chi(s)}{s-s_0}\right)ds=0
\end{equation}

Then:

\begin{equation}
PV \int _{\infty} ^{-\infty} \left(\frac{\chi(s)}{s-s_0}\right)ds= j\pi\left(\chi(s_0)\right)
\end{equation}

Splitting real and imaginary parts:

\begin{equation}
\begin{array}{c}
PV \int _{\infty} ^{-\infty} \left(\frac{\chi^{'}(x)}{x-s_0}\right)dx= \pi\left(\chi^{"}(s_0)\right)\\
-j PV \int _{\infty} ^{-\infty} \left(\frac{\chi^{"}(x)}{x-s_0}\right)dx=j\pi\left(\chi^{'}(s_0)\right)
\end{array}
\end{equation}

Results:

\begin{equation}
\begin{array}{c}
\chi^{"}(s_0)=\frac{1}{\pi}PV \int _{\infty} ^{-\infty} \left(\frac{\chi^{'}(x)}{x-s_0}\right)dx\\
\chi^{'}(s_0)=-\frac{1}{\pi} PV \int _{\infty} ^{-\infty} \left(\frac{\chi^{"}(x)}{x-s_0}\right)dx
\end{array}
\end{equation}

If $s_0=j\omega$, $x=j\xi$, and changing the integral limits the K-K relations can be obtained:
 
\begin{equation}
\begin{array}{c}
\chi^{"}(\omega)=-\frac{1}{\pi}PV \int _{-\infty} ^{\infty} \left(\frac{\chi^{'}(\xi)}{\xi-\omega}\right)d\xi\\
\chi^{'}(\omega)=\frac{1}{\pi} PV \int _{-\infty} ^{\infty} \left(\frac{\chi^{"}(\xi)}{\xi-\omega}\right)d\xi
\label{Relaciones de K K}
\end{array}
\end{equation}

where the integral of the eqn. \ref{Relaciones de K K} is the well known Hilbert transform

\appendixx{Susceptibility of Ferrites}

As a reference the magnetic susceptibility function that can be used in the \textit{s}-complex plane $\chi(s)$ of the MnZn and NiZn soft ferrite can be computed with the following function \cite{Fano20111708}:

\begin{equation}
\chi(s)=\frac{s^3a+s^2b+cs+d}{(s^2 +\beta_1 s+\omega_d^2)(s^2+ s\frac{2\omega_s\alpha}{(1+\alpha^2)}+\frac{\omega_s^2}{(1+\alpha^2)})}\\
\label{eqn:chi}
\end{equation}

where:

\begin{equation}
\begin{array}{c}
a=\frac{(\alpha\omega_s\chi_{s0})}{1+\alpha^2}\\
b=\frac{(1+\alpha^2)\omega_d^2\chi_{d0}+\chi_{s0}\omega_s^2+\alpha\omega_s\chi_{s0}\beta_1}{1+\alpha^2}\\
c=\frac{2\omega_s\alpha \omega_d^2\chi_{d0}+\chi_{s0}\omega_s^2\beta_1+\alpha\omega_s\chi_{s0}\omega_d^2}{(1+\alpha^2)}\\
d=\frac{\omega_s^2\omega_d^2\chi_{d0}+\chi_{s0}\omega_s^2\omega_d^2}{(1+\alpha^2)}
\end{array}
\label{constantes}
\end{equation}

The expressions of the magnetic suseptibility $\chi(s)$ for MnZn and NiZn ferrites have been obtained by Fano et al \cite{Fano20111708}:

\begin{equation}
\chi(s)=\frac{s^34.3951\cdot 10^9+s^28.3019\cdot 10^{16}+3.5404\cdot 10^{23}s+3.0857\cdot 10^{29}}{(s^2 +9.3\cdot 10^6 s+6.25\cdot 10^{12})(s^2+ s3.0086\cdot 10^{13}+5.9706\cdot 10^{26})}\\
\label{chi MnZn}
\end{equation}

\begin{equation}
\chi(s)=\frac{s^37.7202\cdot 10^9+s^28.3570\cdot 10^{16}+2.9710\cdot 10^{23}s+1.7749\cdot 10^{29}}{(s^2 +3.5\cdot 10^6 s+7.84\cdot 10^{12})(s^2+ s1.5030\cdot 10^{16}+5.6481\cdot 10^{31})}\\
\label{chi NiZn}
\end{equation}

\bibliographystyle{ieeetr}
\bibliography{referenciasV7}

\begin{thebibliography}{10}

\bibitem{Wang}
L.~J. Wang, ``Causal all-pass filters and kramers-kronig relations,'' {\em
  Optics Communications}, vol.~213, no.~1-3, pp.~27 -- 32, 2002.

\bibitem{Landau}
L.~D. Landau and E.~M. Lifchitz, {\em Electrodynamics of continuos Media}.
\newblock Addison Wesley, 1981.

\bibitem{Toll}
J.~S. Toll, ``Causality and the dispersion relation: Logical foundations,''
  {\em Phys. Rev.}, vol.~104, pp.~1760--1770, Dec 1956.

\bibitem{Nussenzveig}
H.~M. Nussenzveig, {\em Causality and dispersion relations / H.M. Nussenzveig}.
\newblock Academic Press, New York :, 1972.

\bibitem{Fano1}
W.~G. Fano, S.~Boggi, and A.~C. Razzitte, ``Causality study and numerical
  response of the magnetic permeability as a function of the frequency of
  ferrites using kramers kronig relations,'' {\em Physica B}, vol.~403,
  pp.~526--530, March 2008.

\bibitem{Tsutaoka}
T.~Tsutaoka, ``Frequency dispersion of complex permeability in mn–zn and
  ni–zn spinel ferrites and their composite materials,'' {\em Journal of
  Applied Physics}, vol.~93, pp.~2789--2796, March 2003.

\bibitem{Fano20111708}
W.~G. Fano, S.~Boggi, and A.~C. Razzitte, ``Magnetic susceptibility of mnzn and
  nizn soft ferrites using laplace transform and the routh - hurwitz
  criterion,'' {\em Journal of Magnetism and Magnetic Materials}, vol.~323,
  no.~12, pp.~1708 -- 1711, 2011.

\bibitem{Geyi}
W.~Geyi, {\em Foundations of Applied Electrodynamics}.
\newblock John Wiley, 2010.

\bibitem{Bladel}
J.~Van~Bladel, {\em Electromagnetic Fields}.
\newblock John Wiley, 2007.

\bibitem{Wunch}
D.~A. Wunch, {\em Complex Variables with Applications}.
\newblock Addison-Wesley Company Inc., 1994.

\end{thebibliography}






\end{document}